\documentclass[a4paper,11pt]{article}
\usepackage{pos}

\renewcommand{\Re}{\mathrm{Re}\,}
\newcommand{\Tr}{\mathrm{Tr}\,}

\usepackage{braket}

\title{Test of a two-level algorithm for the glueball spectrum in $SU(N_c)$ Yang-Mills theory}

\author*[a,b]{Andrea Falzetti}
\author[a,b]{Mauro Lucio Papinutto}
\author[a,b]{Francesco Scardino}

\affiliation[a]{Physics department, Sapienza university, Piazzale A. Moro 2, Roma, I-00185, Italy}

\affiliation[b]{Physics department, INFN Roma1, Piazzale A. Moro 2, Roma, I-00185, Italy}

\emailAdd{andrea.falzetti@uniroma1.it}
\emailAdd{mauro.papinutto@roma1.infn.it}
\emailAdd{francesco.scardino@uniroma1.it}

\abstract{We present preliminary results obtained using a new code for $SU (N_c)$ Yang-Mills theory which
performs a 2-level sampling of glueball correlators obtained from a suitably chosen basis of (APE)
smeared and unsmeared operators. The code builds loop operators of any shape and length and
classifies them according to the irreducible representations of the cubic group. We report on the
performances of the algorithm and on the computation of the first low-lying glueball states choosing
$N_c = 3$ as a reference to compare our results with the literature.}

\FullConference{The 41st International Symposium on Lattice Field Theory (LATTICE2024)\\
 28 July - 3 August 2024\\
Liverpool, UK\\}

\begin{document}
\maketitle
\section{Introduction}
The low-lying spectrum of SU($N_c$) Yang-Mills (YM) theory is populated by glueball states \textit{i.e.} bound states composed only of gluons \cite{fritzsch2002currentalgebraquarkselse}. A comprehensive understanding of the glueball spectrum could provide insights into the confinement mechanism and its relation to the mass gap problem. Moreover, any proposed solution of Yang-Mills (YM) theory must reproduce its infrared phenomenology.\\
Physical particles of YM theory are created by gauge invariant composite operators and correspond to glueballs. Experimental efforts to detect a glueball have not managed to produce direct observational evidence of their existence yet \cite{Crede_2009,Ochs_2013}.\\
Though YM perturbation theory is successfully applied in the high-energy regime thanks to the asymptotic freedom, the appearance in YM theory of the renormalization-group (RG) invariant mass scale $\Lambda_{YM}$, which is not analytic in the gauge coupling $g_{YM}$, implies that perturbation theory is unable to compute physical quantities -- such as the glueball masses -- that are proportional to $\Lambda_{YM}$.\\
Lattice discretization is thus an appropriate choice for these types of problems, as it offers a non-perturbative method that is solidly grounded in first principles.\\
For these reasons, the preferred method to study the glueball spectrum is to extract the masses from the long-distance behaviour of correlators computed on the lattice using a Markov Chain Monte Carlo (MCMC) approach.\\ 
However, this approach poses significant numerical challenges; this is because Euclidean correlators vanish exponentially with the time separation between sink and source while the uncertainty remains constant. Thus, the signal-to-noise ratio decreases exponentially in the long-distance limit \cite{parisi_strategy_1984,Lepage:1989hd}.\\
Several numerical techniques have been designed to mitigate this problem, either by increasing the operator's overlap with the lightest states in order to improve the signal (\textit{e.g.} smearing) \cite{apeSm,HypSm} or by mitigating the statistical noise in the correlator's measurements such as the multi-hit \cite{multihit} and multilevel \cite{meyer_locality_2003,meyer_yang-mills_2004} algorithms.\\
In particular, the multilevel algorithm exploits the locality of the Wilson plaquette action to factorize the path integral into contributions coming from spacetime regions separated by a fixed boundary. By performing repeated submeasurements of the operators in these subregions, we can estimate the correlator by computing a nested average of our measurements, achieving an exponential error suppression in time separation.\\
In this contribution, we discuss the structure of the algorithm and present preliminary results on the error-suppression properties of this approach for $N_c = 3$ on 2 different lattices. We also provide estimates of the lowest-lying glueball states in order to validate our algorithm by comparing our results with the literature.

\section{The Multilevel algorithm}\label{Sec:Multi}
The Wilson plaquette action \cite{montvay_quantum_1994} 
\begin{equation}
    \label{WilsPlaqAction}
S[U] = \sum_p \frac{1}{N_c}\Re\Tr\left(1-U_p\right)
\end{equation}
is local \textit{i.e.}  each link variable $U_\mu(x)$ is only coupled to the links that live in a finite radius from $x$. This is more easily understood by examining the local contribution of each link to the action:
\begin{equation}
S(U) = - \frac{1}{N_c}\Re\Tr \left(U_\mu(x)\cdot V_\mu(x)^\dagger\right)
\end{equation}
Where $V$ is the sum of all the staples that one can build around the link $U_\mu(x)$. \\


If we divide our lattice into two distinct regions, $\Lambda_1$ and $\Lambda_2$, separated by a fixed boundaries $\partial \Lambda$ in such a way that for each $U_\mu(x)$ the corresponding $V_\mu(x)$ only involves link variables residing in $\Lambda_1$ or on the boundary\footnote{For the Wilson simple plaquette action we can obtain this by choosing as boundary the spatial link variables residing on two separate timeslices. For improved actions involving bigger Wilson loops it can be necessary to widen the boundaries between the two regions to ensure separation between the two regions.}, and vice versa for $\Lambda_2$, then the evolution in Monte Carlo time of the $\Lambda_1$ configuration will be completely independent from the $\Lambda_2$ configuration.\\
If we want to compute a matrix of euclidean correlators between two different spatial Wilson loops $\mathcal{O}_i(t_1)$ and $\mathcal{O}_j(t_2)$\footnote{Considering that we want to compute the spectrum we will only be interested in the $p=0$ projection of these operators which amounts to summing the value of the loop over all sites in a timeslice. Thus our operators only depend on the euclidean time variable.} belonging to some operator basis, such that $t_1 \in \Lambda_1$ and $t_2 \in \Lambda_2$ the path integral expression reads:
\begin{align}
    &\left<\mathcal{O}_i(t_1)\mathcal{O}_j(t_2)\right> = \frac{1}{Z}\int_{\Lambda_1 \cup \Lambda_2 \cup \partial \Lambda} d[U] e^{-\beta S[U]} \mathcal{O}_i(t_1)\mathcal{O}_j(t_2) \nonumber\\
    &= \frac{1}{Z}\int_{\partial \Lambda} d[U]_{\partial \Lambda}\, e^{-\beta S[U]_{\partial \Lambda}}\left(\int_{\Lambda_1} d[U]_{\Lambda_1} \, e^{-\beta S[U]_{\Lambda_1}} \mathcal{O}_i(t_1)\right) \left(\int_{\Lambda_2} d[U]_{\Lambda_2} \, e^{-\beta S[U]_{\Lambda_2}} \mathcal{O}_j(t_2)\right) \nonumber \\
    &\equiv \left<\left<\mathcal{O}_i(t_1)\right>_{\Lambda_1} \left<\mathcal{O}_j(t_2)\right>_{\Lambda_2}\right>_{\partial \Lambda}
\end{align}
The path integral factorizes into two independent path integrals, one for each region, combined into an overall average over the different gauge configurations we can have on the boundary.
 If we translate this into our Monte Carlo averages
\begin{equation}
    \label{MCaverage}
C_{ij}(t,t_0) \equiv \left<\mathcal{O}_i(t)\mathcal{O}_j (t_0)\right> =  \frac{1}{N_{\mathrm{conf}}}  \sum_{\ell = 1}^{N_{\mathrm{conf}}}\mathcal{O}_i(t)([U]_\ell) \mathcal{O}_j(t_0)([U]_\ell) 
\end{equation}
the matrix of correlators becomes a nested average as we explain below.\\
\begin{figure}
    \centering
    \includegraphics[width=0.65\linewidth]{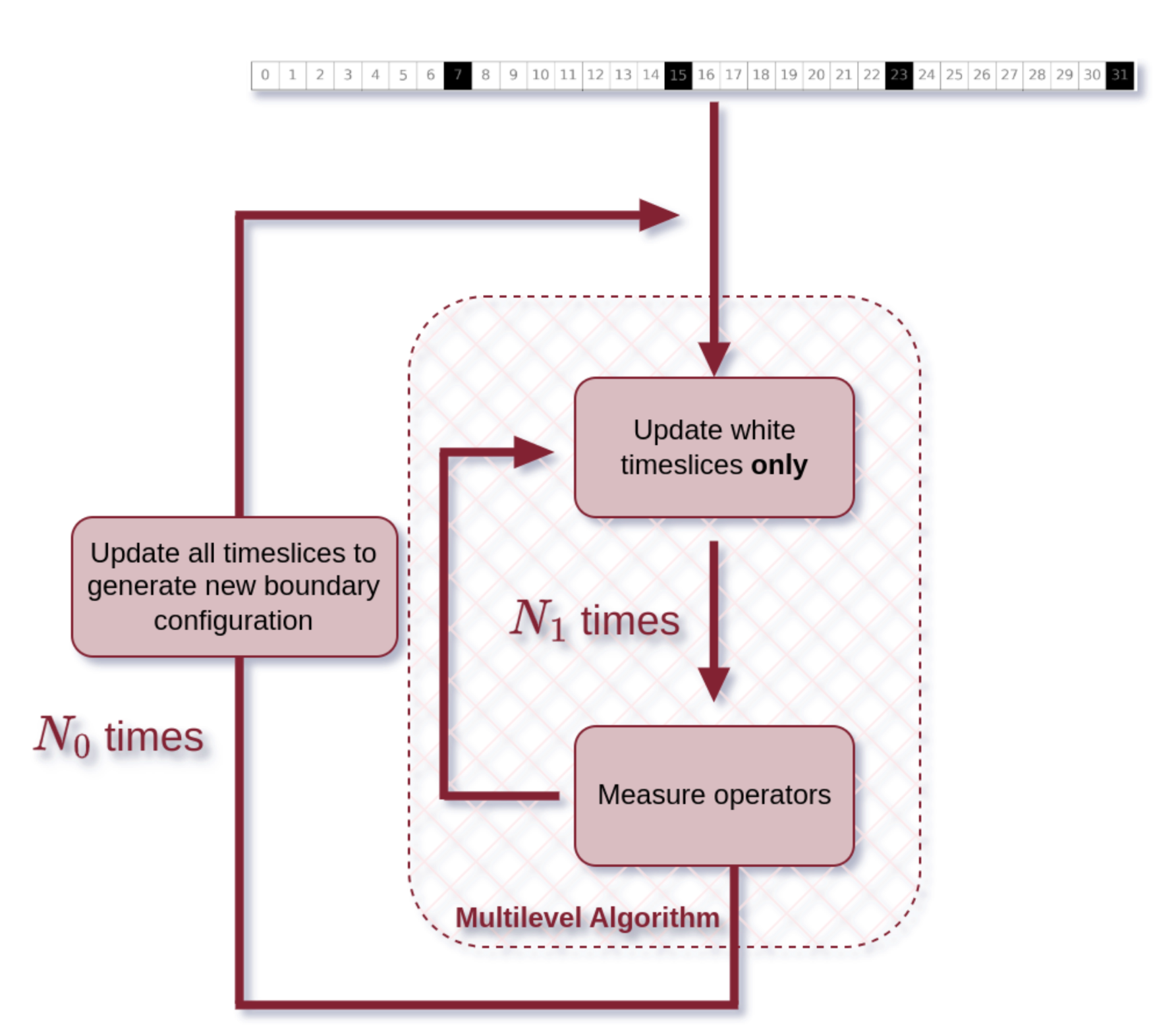}
    \caption{Flow diagram describing the 2-level algorithm used to compute gluonic observables. The image above is a pictorial representation of the partitioning of the lattice implied by the use of the algorithm.}
    \label{fig:FlowDiag}
\end{figure}
On the lattice this is obtained by performing a cycle of sub-updates and sub-measurements as depicted in the diagram in figure \ref{fig:FlowDiag}. Thus, for each boundary configuration $[U]_{\partial \Lambda}$ we generate $N_1$ sub-configurations of the two regions $\Lambda_1$ and $\Lambda_2$ on which we measure and average our operators. We repeat this process on $N_0$ different boundary configurations in order to accurately sample the whole phase space. The result is a dataset composed of $N_0\cdot N_1$ measurements for each operator $\mathcal{O}_i^{(b,\ell)}(t)$ with $b = 1, ..., N_0$ and $\ell= 1, ..., N_1$; the correlator can now be measured on the lattice with the nested average:
\begin{equation}
    \label{NestedSumMultilevel}
C_{ij}(t,t_0) \equiv \left<\mathcal{O}_i(t) \mathcal{O}_j(t_0)\right> = \frac{1}{N_0 \cdot N_1^2} \cdot \sum_{b = 1}^{N_0} \left(\sum_{\ell = 1}^{N_1} \mathcal{O}_i^{(b,\ell)}(t)\right) \left(\sum_{\kappa = 1}^{N_1} \mathcal{O}_j^{(b,\kappa)}(t_0)\right)
\end{equation}
Where we assume that $t$ and $t_0$ are separated by at least one frozen timeslice.\\
Effectively, we only performed $N_0\cdot N_1$ updates. However, due to the factorization of the sums in equation \eqref{NestedSumMultilevel} we have produced $N_0 \cdot N_1^2$ measurements of the correlator. This implies an error suppression of order $\frac{1}{N_1}$ rather than $\frac{1}{\sqrt{N_1}}$ as one would naively expect when performing $N_0\cdot N_1$ updates.\\ \\
\subsection{The algorithm in practice}
When implementing the algorithm on the lattice, some considerations must be made to ensure its effectiveness.\\
Firstly, the algorithm relies on the spacetime factorization in the path-integral, as described in the preceding paragraph; thus it can only be applied to $2$-point correlators where the sink and the source are located in different regions, separated by one, or more, frozen timeslices. Furthermore, the fixed boundaries enclosing each region turn out to be a consistent noise source for the correlators, thus the best performance is obtained when sink and source are placed as far away from the boundaries as possible \cite{Schaefer_err}.\\
In order to extract the spectrum of the theory we need an accurate estimate of the correlators in the large time-separation limit, when the effective mass (or the GEVP mass if one is using a variational technique) approaches a plateaux. These considerations should always be kept in mind when choosing the spacing between the different multilevel boundaries since we want the performance of the algorithm to be as effective as possible in the time region where the noise becomes relevant, so that the plateau becomes longer and more easily identifiable.\\

Furthermore, if we only rely on measurements obtained through multilevel averaging, we have no hope of ever measuring the correlator when $\left|t_1-t_0\right| < 2a$ since in this case source and sink will always be in the same region or on the boundaries. This means we won't be able to compute the 2-point correlator $C(\Delta t)$ for $\Delta t = 0,a$\footnote{And its symmetric twins at $\Delta t = L0, L0 -a $ where $L0$ is the temporal extension of the lattice.}. This is problematic, especially when trying to combine this algorithm with the GEVP technique, as one might want to use those points as the pivot $C(t_0)$. 
To solve this issue, we employ a "multilevel agnostic" approach as was done in \cite{Schaefer}: for each couple of values $t_0, t_1$, if sink and source are in two different regions we utilize the multilevel nested averaging exploiting the factorization, if they are not we simply treat them as independent measurements of the operator and measure the correlator in the naive way. After assigning an error to each $C(t_0, t_1)$ we compute the 2-point correlator $C(\Delta t)$ with a weighted average over all $C(t_0,t_1)$ such that $|t_1-t_0| = \Delta t$ using the inverse variances on the correlator as weights.\\
\section{Operator basis}
\begin{figure}[h!]
    \centering
    \begin{minipage}{0.47\linewidth}
        \centering
        \begin{tabular}{|c|c|}
        \hline
        $J^{PC}$ & $N_{\mathrm{ops}}$ \\ 
        \hline
        $A_1^{++}$ & $7$ \\ 
        $A_2^{++}$ & $2$ \\ 
        $E^{++}$ & $18$ \\ 
        $T_1^{++}$ & $3$ \\ 
        $T_2^{++}$ & $6$ \\ 
        $T_1^{-+}$ & $9$ \\ 
        $T_2^{-+}$ & $12$ \\ 
        \hline
        \end{tabular}
        \caption{Irreducible representations content of our operator basis.}
        \label{tab:operatorsRep}
    \end{minipage}%
    \hspace{0.04\linewidth} 
    \begin{minipage}{0.48\linewidth}
        \centering
        \includegraphics[width=\linewidth]{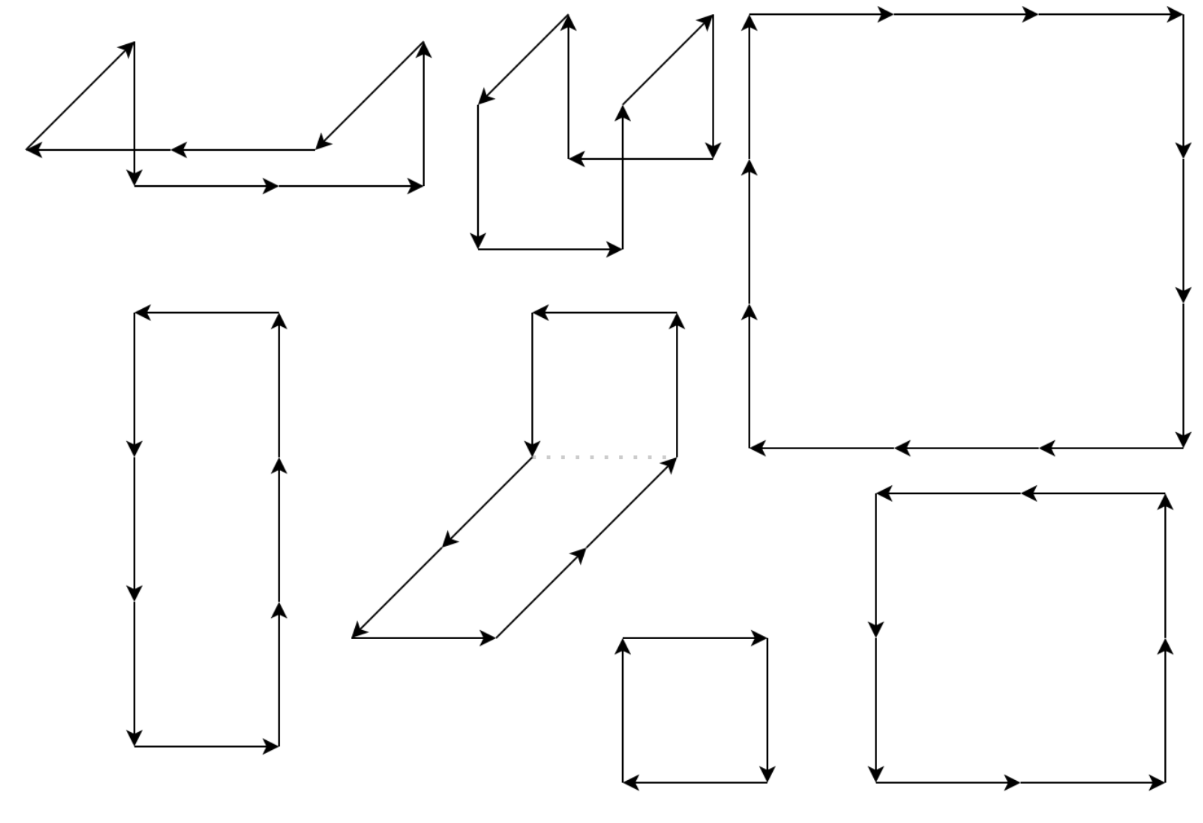}
        \caption{Pictorial representation of the Wilson loops composing our operator basis.}
        \label{fig:loops}
    \end{minipage}
\end{figure}

We have chosen the operator basis to be a selection of Wilson loops computed at several smearing levels. We picked our operators in order to maximize the number of independent operators in the $A_1^{++}$ and $E^{++}$ representations, containing the Spin $0^{++}$ and Spin $2^{++}$ glueballs respectively, which are the most studied in literature, while keeping the computational cost of the simulation at a reasonable level.\\
In our code, once the user selects the loop shapes to be included in the measurements, the program automatically applies all cubic group rotations to the selected shapes and then builds the correct linear combinations which transform according to the irreducible representations of $O_h$.\\
The smearing procedure is implemented according to the APE scheme \cite{apeSm}, and tuned differently with respect to the lattice size as to avoid back-propagation of the smeared links along the periodic boundaries.
The decomposition of the operator basis in terms of irreducible representations of the cubic group is depicted in figure \ref{tab:operatorsRep}, while figure \ref{fig:loops} contains a pictorial representation of all the loop shapes we have measured.

\section{Results}
\begin{figure}[]
    \centering
    \begin{minipage}{0.45\linewidth}
        \centering
        \includegraphics[width = .95\linewidth]{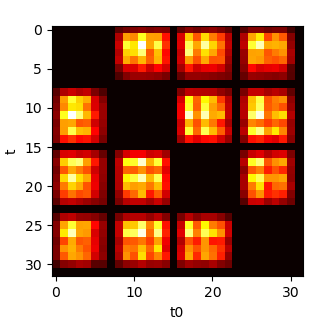}
    \end{minipage}%
    \hspace{0.06\linewidth} 
    \begin{minipage}{0.45\linewidth}
        \centering
        \includegraphics[width=\linewidth]{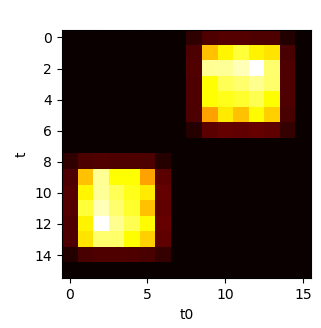}
    \end{minipage}
    \caption{Heat map of the weights in the source position average for the two different lattices. The weights are proportional to $\sigma^{-2}(C(t,t_0))$. As we can see the geometry of the multilevel setup is clearly visible in the weights behaviour.} 
    \label{heatmap}
\end{figure}

The simulations are performed on two different $4$-dimensional lattices with periodic boundary conditions, the details of the lattices with the amount of statistics collected on each are depicted in table \ref{tab:Lattices}. 
\begin{table}[h!]
\centering
\begin{tabular}{|cccccccc|}
\hline
\textbf{Lattice n.} & \textbf{$L_0$} & \textbf{$L_i$} & \textbf{$\beta$} & \textbf{$\Delta_{\mathrm{Multilevel}}$} & \textbf{$N_0$} & \textbf{$N_1$} & \textbf{$a[fm]$} \\ \hline
1 & 32 & 24 & 6.2 & 8 & 160 & 500 &  0.068 \\
2 & 16 & 16 & 5.95 & 8 & 160 & 500 &  0.102\\\hline
\end{tabular}
\caption{Summary of the lattice setups used in this work. The meaning and relevance of the $\Delta_{\mathrm{Multilevel}}$, $N_0$ and $N_1$ parameters are explained in section \ref{Sec:Multi}}
\label{tab:Lattices}
\end{table}
The first indication that the algorithm effectively suppresses the error of the correlator in the $\Delta t >> a$ region is the fact that when we perform the weighted average over the source positions, combinations of $t,t_0$ referring to timeslices in different dynamical regions tend to have a disproportionately large weight as depicted by figures \ref{heatmap}. The heat maps reveal that the analysis strongly favours source-sink combinations that leverage the previously described path-integral factorization.\\ \\

In table \ref{tab:masses} we list the ground states in the symmetry channels for which it was possible to extract a non-ambiguous plateau using the GEVP method.

\begin{table}[h!]
\centering
\begin{tabular}{r|cc|cc|}
\cline{2-5}
 & \multicolumn{2}{c|}{\textbf{\shortstack{Lat. 1 \\ ($\beta = 6.2$)}}} & \multicolumn{2}{c|}{\textbf{\shortstack{Lat. 2 \\ ($\beta = 5.95$)}}} \\ \cline{2-5} 
 & $a\cdot m_0$ & \cite{MeyerGlu} & $a\cdot m_0$ & \cite{Mondal} \\ \hline
\multicolumn{1}{|l|}{$A_1^{++}$} & 0.525 (12) & 0.5197 (51) & 0.752 (4) & 0.7510 (15) \\
\multicolumn{1}{|l|}{$E^{++}$} & 0.776 (25) & 0.7784 (79) & 0.951 (40) & 0.938 (17) \\ \hline
\end{tabular}
\caption{Ground state masses of glueballs in the scalar ($A_1^{++}$) and spin 2 channel ($E^{++}$). In the first columns of each lattice we show our results while in the second columns we compare with \cite{MeyerGlu} and \cite{Mondal} respectively}
\label{tab:masses}
\end{table}

\begin{figure}[]
    \centering
    \includegraphics[width=.5\linewidth]{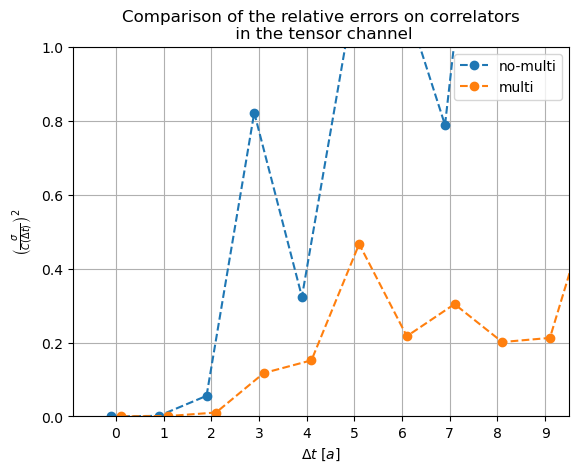}
    \caption{Comparison between the squared relative variance $\left(\frac{\sigma^2}{C(\Delta t)}\right)^2$ in runs with or without multilevel at equivalent statistics}
    \label{comparison_multi}
\end{figure}

We start by checking that the algorithm achieves error suppression on the correlator entries, in figure \ref{comparison_multi} we compare the squared relative variance on correlators in the $E^++$ channel with data from an independent run with equivalent statistics computed without using multilevel techniques. The data shows substantial equivalence for values of $\frac{\Delta t}{a} < 2$ as there is no way to have a $(t,t+1)$ combination with both timeslices sitting in different dynamical regions, then as the multilevel correlators start dominating the average the data shows a much slower signal-to-noise ratio decay. 

\begin{figure}[]
    \centering
    \includegraphics[width=.6\linewidth]{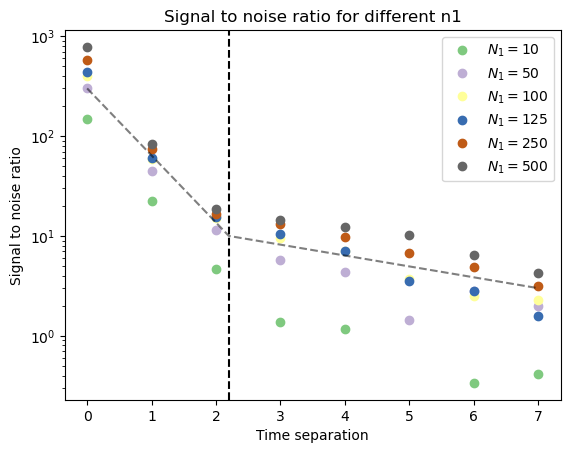}
    \caption{Exponential decay of the signal to noise ratio for correlators in the $A_1^{++}$ channel for different values of $N_1$. The dashed line is used to illustrate the change in the rate of exponential decay when multi-level averaging takes effect.}
    \label{n1_comparison}
\end{figure}

\begin{figure}[]
    \centering
    \includegraphics[width=.6\linewidth]{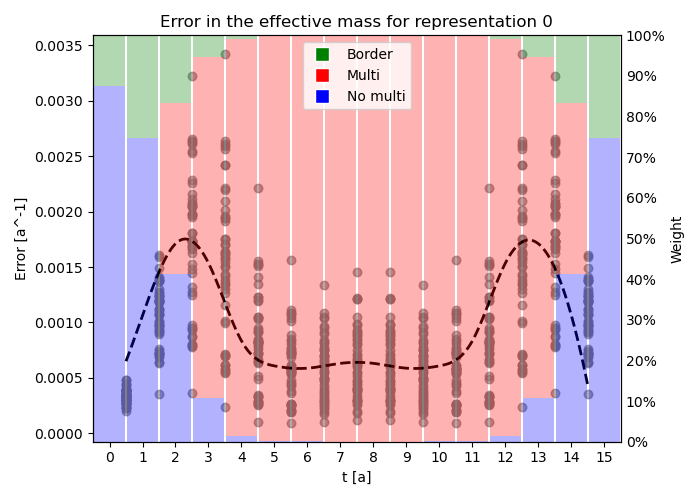}
    \caption{Error in the effective mass for the correlators in the $A_1^{++}$ representation overlapped with a bar plot of the average weight contribution of multilevel data. The points at each $t$ represent the errors on the effective mass of the diagonal entries of the matrix of correlators. The bar chart is generated from averaging the weights of all entries of the correlator in the moving-sink average and computing the percentage contribution of each type of $t,t_0$ combination: "Border" represents correlators for which either $t$ or $t_0$ or both stand on the frozen timeslices, "No multi" indicates those cases where $t$ and $t_0$ are in the same active region, "Multi" indicates those combinations where $t$ and $t_0$ are separated by a frozen timeslice and thus exploit the path-integral factorization. The dashed line is a smoothed spline interpolation of the average data to guide the eye.}
    \label{fig:effmass}
\end{figure}

Next we would like to investigate the performance's dependence on the $N_1$ parameter which regulates the amount of submeasurements performed at each fixed boundary configuration. To do this we have taken all correlators of the type $C_{ii}(t)$\footnote{The choice to only use the diagonal entries is dictated by readability and the fact that these tend to be less noisy overall} in the $A_1^{++}$ channel and plotted the average signal to noise ratio at different values of $\Delta t$. This quantity is expected to vanish exponentially for large times, from the usual Lepage-Parisi argument \cite{parisi_strategy_1984,Lepage:1989hd}. Looking at the data in figure \ref{n1_comparison} we see that after the first few timeslices, when the multilevel data start dominating the average, the decay behaviour of the signal-to-noise ratio slows down dramatically and a new exponential regime is observed. 
Finally, we want to check the algorithm's performance in improving the mass measurements. The statistics of the ensembles are not large enough to directly confront the GEVP eigenvalues results. If we compute the effective mass for all correlators in the $A_1^{++}$ representation and analyse the error in $m_{eff}(t)$ we can see in Figure \ref{fig:effmass} that as we move towards a bigger time separation, the multilevel data become predominant in the average and the error is suppressed.

\end{document}